\documentclass[aps,pre,showpacs,preprint]{revtex4}
\usepackage{amsmath}
\usepackage{amsfonts}
\usepackage{graphicx}
\usepackage{overpic}
\usepackage{psfrag}
\usepackage{color}

\begin{document}

\title{Structure and stability of double emulsions}
\author{J.~Guzowski}
\affiliation{Max--Planck--Institut f\"ur Metallforschung,
Heisenbergstr.\ 3, 70569 Stuttgart, Germany\\
and\\
Institut f\"ur Theoretische und Angewandte Physik, Universit\"at Stuttgart,
 Pfaffenwaldring 57, 70569 Stuttgart, Germany}

\author{P.~Garstecki}
\affiliation{Institute of Physical Chemistry, Polish Academy of Science,\\
ul. Kasprzaka 44/52, 01-224 Warsaw, Poland}

\author{P.~Korczyk}
\affiliation{Institute of Physical Chemistry, Polish Academy of Science,\\
ul. Kasprzaka 44/52, 01-224 Warsaw, Poland}

\date{\today}

\begin{abstract}
We propose diagrams representing the equilibrium morphologies of two immiscible liquid droplets brought into contact. We study the dependence of the shape of the droplets on the surface tensions and ratio of volumes. We study theoretically and experimentally the regimes of the parameters corresponding to complete engulfing (one of the droplets completely absorbed by the other one), non-engulfing (in which the droplets remain separated) and partial-engulfing (intermediate configuration in which all three interfaces are present). We specify the values of surface tensions corresponding to a Janus droplet, i.e., a perfectly spherical droplet composed of two sub-volumes occupied by two different phases and separated by a curved interface. We further calculate and experimentally verify a morphological transition between the states with positive and negative curvature of this interface depending on the ratio of volumes and on the equilibrium contact angle. 
\end{abstract}


\maketitle

\section{Introduction.}
When droplets of two different immiscible liquids immersed in the third host liquid are brought into contact, there are three possible equilibrium topologies which can form. Depending on the surface tensions between the three liquid phases the droplets can remain separated by the host phase, one of the phases can completely absorb the other one, or a dumbbell-like doublet can form, in which all three possible interfaces are present~\cite{Torza1969}. In this work we study in detail the stability diagram of all the morphologies with the focus on the so-called Janus states in which the doublet forms a perfect sphere, in analogy to colloidal Janus particles (whose surface is divided into two chemically varying regions - usually hemispheres). In such a case the spherical drop has two domains occupied each by a different phase. We investigate the sign of the curvature of the interface between these two phases as a function of the contact angle and the ratio of volumes. Experimentally, the latter can be controlled by using a microfluidic device, in which the pairs of droplets of desired volumes can be manufactured in a reproducible manner by adjusting the relative flow speeds of the different species~\cite{Xu2005}. 

Multiple droplets have already found a number of applications in medicine and industry. Encapsulation of active substances in monodisperse particles enhances the control over the profile of release~\cite{Xu2009}. It has also been demonstrated that encapsulation of cells from the host tissue enhances their viability~\cite{Schmidt2008} and that it is possible to encapsulate microorganisms~\cite{Choi2007}, enzymes~\cite{Um2008} or blood proteins~\cite{Liu2008}. The viability of encapsulated cells may depend on diffusion of media, stimuli and waste products, and on the geometry of the encapsulant. Further, the rate of release of encapsulated molecules can be controlled via tuning of the architecture of multiple emulsions (i.e. number of shells, number, composition and size of inner droplets).

On the other hand ensembles of Janus droplets can be further processed (e.g. by photo-curing) in order to fabricate colloidal Janus particles with a prescribed internal structure~\cite{Panacci2008} as well as a prescribed surface composition. A precise control of the surface ratio of a Janus particle is also a big advantage in designing self-propelling particles.
Such particles, due to their surface activity, produce gradients of concentration of a dissolved species in the surrounding, which leads to a directed motion. Particularly, the speed of the particle depends crucially on the size and the activity of the chemically active region~\cite{Golestanian2007}. On the other hand Janus particles self-assemble into clusters the size of which depends on the surface composition of the particles~\cite{Nie2006}. Finally, assemblies of transparent Janus droplets can be used to fabricate micro-lens arrays, with potential application in microoptics~\cite{Yabu2005}.

\section{Theory}

\begin{figure}[ht]
 \psfragscanon
  \psfrag{0}[c][c][1]{$0$}
  \psfrag{A}[c][c][1]{$A$}
  \psfrag{B}[c][c][1]{$B$}
  \psfrag{t0}[c][c][1]{$\theta_0$}
  \psfrag{ta}[c][c][1]{$\theta_A$}
  \psfrag{tb}[c][c][1]{$\theta_B$}
  \psfrag{r}[c][c][1]{$R$}
  \psfrag{ra}[c][c][1]{$R_A$}
  \psfrag{rb}[c][c][1]{$R_B$}
  \psfrag{gab}[c][c][1]{$\gamma_{AB}$}
  \psfrag{ga}[c][c][1]{$\gamma_{0A}$}
  \psfrag{gb}[c][c][1]{$\gamma_{0B}$}
\vspace*{0.7cm}
\hspace*{3.5cm}
 \begin{overpic}[width=0.6\textwidth]{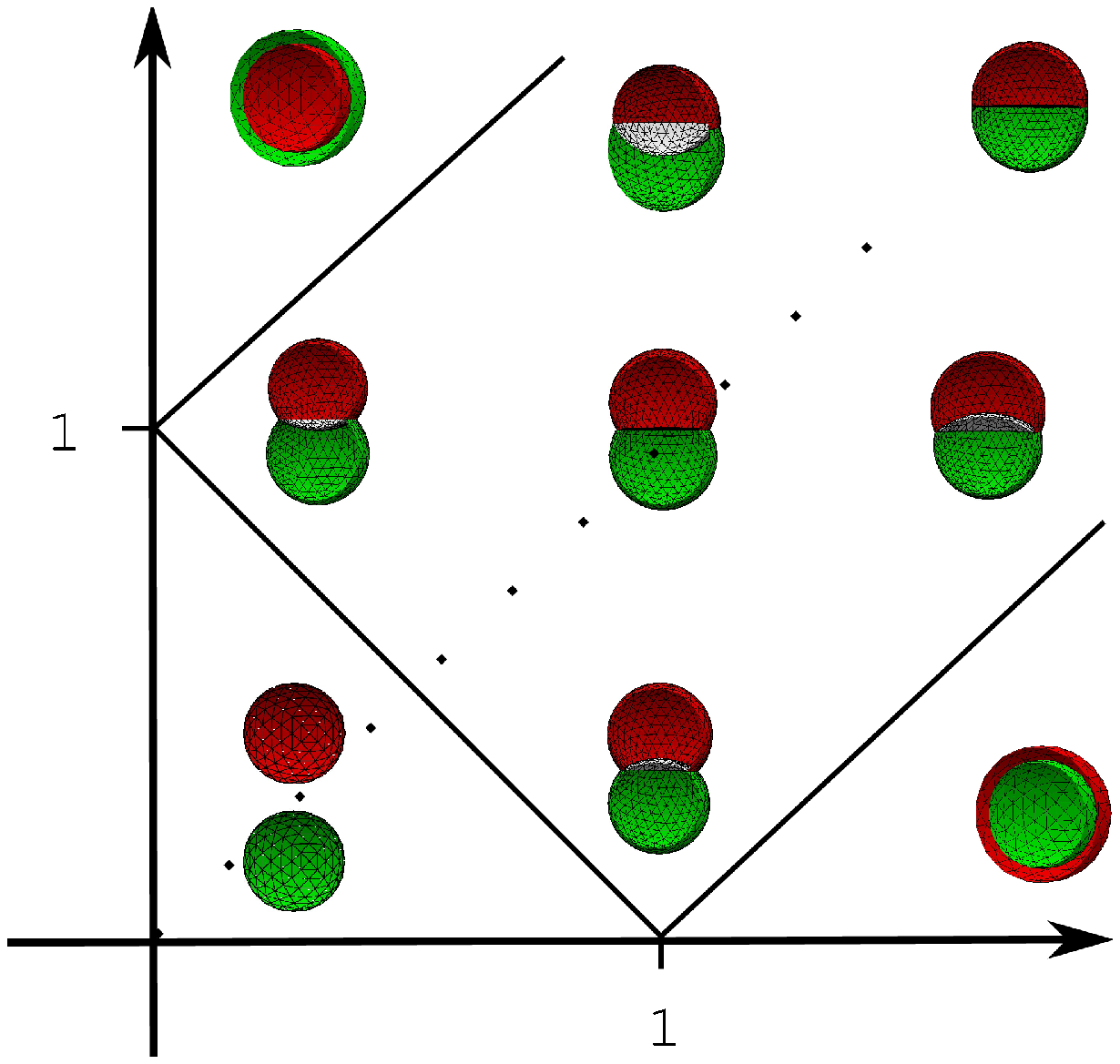}
  \put(73,15){engulf.}
  \put(33,15){non-engulf.}
  \put(50,42){partially-engulf.}
  \put(15,73){engulf.}
  \put(80,73){Janus}
  \put(95,5){$\gamma_{0A}/\gamma_{AB}$}
  \put(5,100){$\gamma_{0B}/\gamma_{AB}$}
  \put(-45,10){\includegraphics[width=0.2\textwidth]{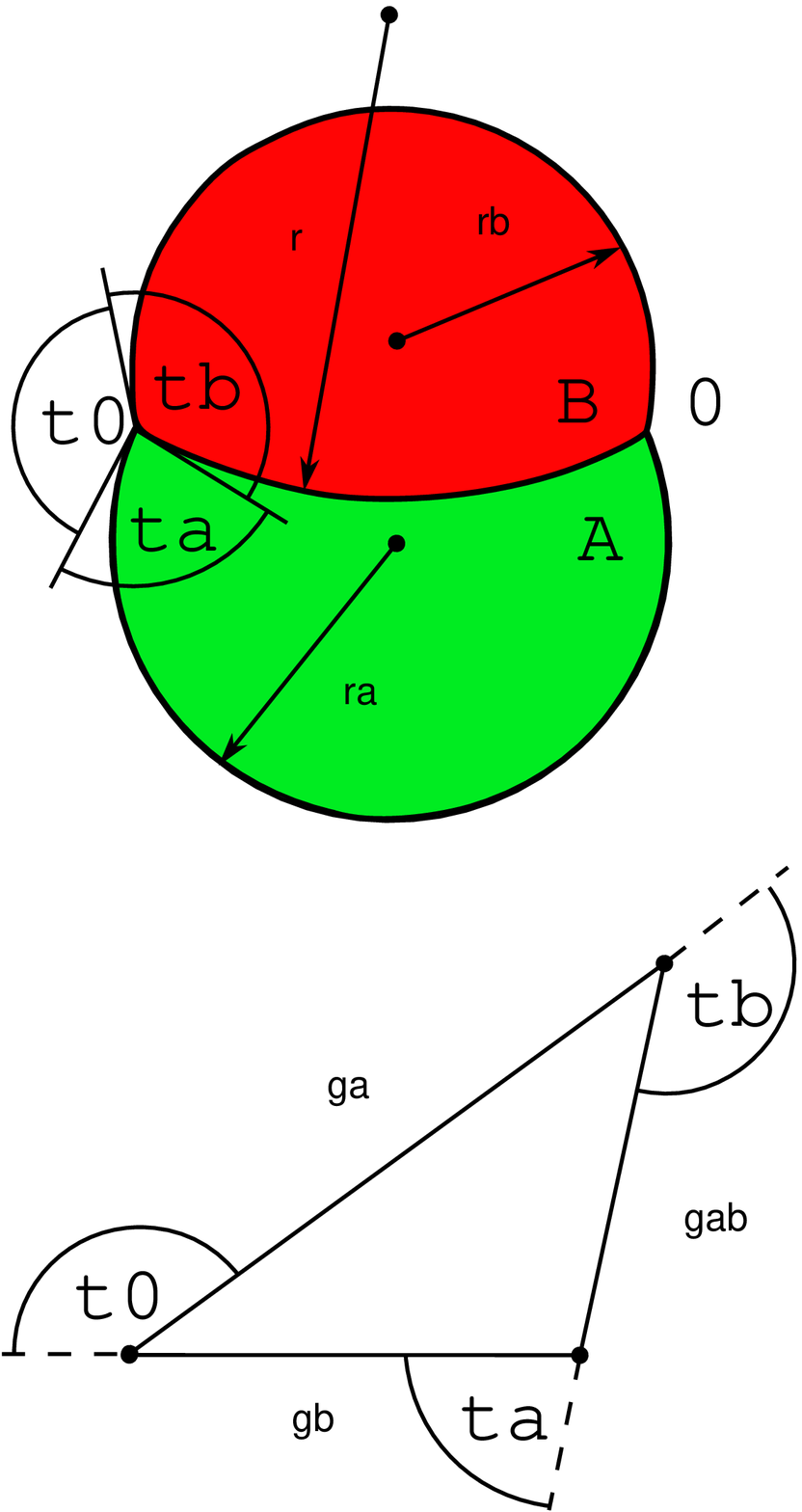}}
\psfragscanoff
 \end{overpic}
 \caption{Left: sketch of the partially engulfing configuration with the phases $A$ and $B$ surrounded by the phase $0$ and the Neumann's triangle whose sides have lengths proportional to the surface tensions. Right: the diagram representing possible morphologies formed by the phases $A$ (green) and $B$ (red) in the case of equal droplet volumes $V_A=V_B$ (the snapshots are from finite-element surface free energy minimization obtained with the open-source Surface Evolver~\cite{Brakke}). The dotted line corresponds to the condition $\theta_B=\theta_A$.}
 \label{fig:diagram}
\end{figure}

We consider two immiscible liquids $A$ and $B$ of volumes $V_A$ and $V_B$, respectively, in contact with each other and both immersed in the third host liquid referred to with index $0$. In such a system three interfaces can form, characterized by three different surface tensions $\gamma_{0A}$, $\gamma_{0B}$ and $\gamma_{AB}$. In mechanical equilibrium each of the interfaces is a section of a sphere and therefore the equilibrium shape of the droplets is fully determined by the contact angles  $\theta_A$ and $\theta_B$ between the interfaces at the three-phase contact line and by the ratio of the liquid volumes $V_B/V_A$. Mechanical equilibrium of the contact line demands that the contact angles are related to the three surface tensions $\gamma_{0A}$, $\gamma_{0B}$ and $\gamma_{AB}$ via the set of equations:
\begin{align}
 &\gamma_{AB} + \gamma_{0B}\cos\theta_B + \gamma_{0A}\cos\theta_A=0,\label{eq:neumann1}\\
 &\gamma_{AB}\cos\theta_B + \gamma_{0B} + \gamma_{0A}\cos\theta_0=0,\\
 &\gamma_{AB}\cos\theta_A + \gamma_{0B}\cos\theta_0 + \gamma_{0A}=0,\label{eq:neumann2}
\end{align}
whose geometrical interpretation is known as the Neumann's triangle (see Fig.\ \ref{fig:diagram}). Only two of the three contact angles are independent, say $\theta_A$ and $\theta_B$, whereas $\theta_0=2\pi-\theta_A-\theta_B$. However, it might also happen that for given surface tensions the set of Eqs.\ (\ref{eq:neumann1})-(\ref{eq:neumann2}) admits no solution for $\theta_A$ and $\theta_B$. Physically, this means that, in equilibrium, there is no three-phase contact line at which the contact angles could be measured. This indicates the change in the topology. The distinction between the various possible topologies can be conveniently done in terms of the so-called spreading coefficients defined as $S_i:=\gamma_{jk}-\gamma_{ik}-\gamma_{ij}$, $\{i,j,k\}=\{0,A,B\}$. In terms of wetting, the spreading coefficient reflects the tendency of a liquid to spread on a solid substrate. In our case the substrate is replaced by an interface between two other liquids. When $S_i>0$ the liquid $i$ spreads on $jk$-interface into a thin film. When $S_i<0$ it rather forms a liquid lens characterized by a finite angle between $ik$ and $ij$-interfaces. Depending on the values of the spreading parameters one can distinguish the three possible morphologies:

\begin{itemize}
\item {\em engulfing}: the phase $A$ is entirely absorbed into the phase $B$ ($S_B>0$) or vice versa ($S_A>0$); 

\item {\em non-engulfing}: droplets of phases $A$ and $B$ are separated by the host phase $0$ ($S_0>0$).

\item {\em partial-engulfing}: droplets of phases $A$ and $B$ touch each other and are both exposed to the host phase $0$. In this case one can solve Eqs.\ (\ref{eq:neumann1})-(\ref{eq:neumann2}) with respect to $\cos\theta_A$ and $\cos\theta_B$ with the result
\begin{align}
 \cos\theta_A&=\dfrac{\gamma_{0B}^2-\gamma_{AB}^2-\gamma_{0A}^2}{2\gamma_{AB}\gamma_{0A}},\label{eq:cosA}\\
 \cos\theta_B&=\dfrac{\gamma_{0A}^2-\gamma_{AB}^2-\gamma_{0B}^2}{2\gamma_{AB}\gamma_{0B}}.\label{eq:cosB}
\end{align}
The mathematical bounds $-1<\cos\theta_{A}<1$ and $-1<\cos\theta_{B}<1$ translate into the conditions $S_A<0$, $S_B<0$, and $S_0<0$.
\end{itemize}

In the following Subsection we focus on the situations of partial-engulfing and study in more detail various possible shapes of a doublet.  

\subsection{Partially-engulfing states}
We start the analysis from Eqs.\ (\ref{eq:cosA}) and (\ref{eq:cosB}). One can distinguish the following characteristic limiting cases:

\begin{itemize}

 \item {\em solid-like phase $A$}: $\gamma_{AB}\rightarrow\gamma_{0A}$ and $\gamma_{0B}/\gamma_{AB}\rightarrow 0$; keeping only the leading terms in $\gamma_{0A}-\gamma_{AB}$ and $\gamma_{0B}$ in Eqs.\ (\ref{eq:cosA}) and (\ref{eq:cosB}) leads to
 \begin{align}
  \cos\theta_A&=-1, \label{eq:solid-a}\\
  \cos\theta_B&=\dfrac{\gamma_{0A}-\gamma_{AB}}{\gamma_{0B}}=-\cos\theta_0 \label{eq:solid-a2}
 \end{align}
 From Eq.\ (\ref{eq:solid-a}) it follows that $\theta_A=\pi$, which means that the droplet of phase $A$ forms a perfect sphere which behaves at the $0B$-interface like a solid particle, i.e., in this limit it does not deform upon changing the surface tensions, but only changes its immersion into the phase $B$, see Fig.\ \ref{fig:solid-a}. Accordingly, Eq.\ (\ref{eq:solid-a2}) expresses a usual Young's law for the contact angle at a solid particle.
\begin{figure}[ht]
\psfragscanon
\psfrag{t0}[c][c][1]{$\theta_0$}
\psfrag{tA}[c][c][1]{$\theta_A=\pi$}
\psfrag{A}[c][c][1]{$A$}
\psfrag{B}[c][c][1]{$B$}
 \begin{overpic}[width=0.5\textwidth]{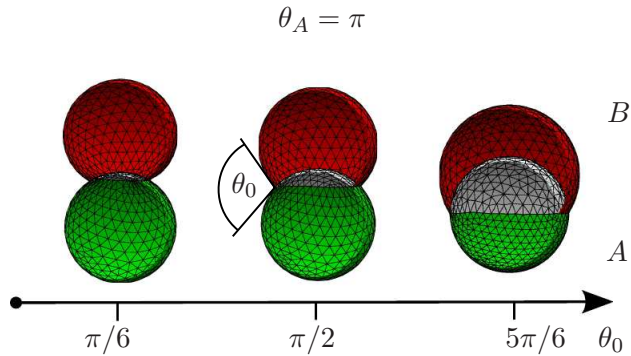}
  \put(95,-4){$\theta_0$}
  \put(12,-4){$\pi/6$}
  \put(45,-4){$\pi/2$}
  \put(80,-4){$5\pi/6$}
 \end{overpic}
\psfragscanoff
 \caption{Configurations of the droplets for various $\theta_0$ in the limiting case $\gamma_{AB}\rightarrow\gamma_{0A}$ and $\gamma_{0B}/\gamma_{AB}\rightarrow 0$, which corresponds to $\theta_A=\pi$.}
 \label{fig:solid-a}
\end{figure}

 \item {\em solid-like phase $B$}: $\gamma_{AB}\rightarrow\gamma_{0B}$ and $\gamma_{0A}/\gamma_{AB}\rightarrow 0$; because the species $A$ and $B$ are equivalent, there is full analogy to the previous case upon interchanging $A$ and $B$;

 \item {\em Janus-like doublet}: $\gamma_{0B}\rightarrow\gamma_{0A}$ and $\gamma_{0A}/\gamma_{AB}\rightarrow \infty$; in this limit one obtains
 \begin{equation}
  \cos\theta_B=\dfrac{\gamma_{0A}-\gamma_{0B}}{\gamma_{AB}}=-\cos\theta_A. \label{eq:janus}
 \end{equation}
 This translates into $\theta_A+\theta_B=\pi=2\pi-\theta_0$, i.e., $\theta_0=\pi$. Physically, this corresponds to a Janus-like state in which the doublet forms a perfect sphere, see Fig.\ \ref{fig:janus};  
\begin{figure}[ht]
\psfragscanon
\psfrag{A}[c][c][1]{$A$}
\psfrag{B}[c][c][1]{$B$}
\psfrag{tb}[c][c][1]{$\theta_B$}
\psfrag{t0}[c][c][1]{$\theta_0=\pi$}
 \begin{overpic}[width=0.5\textwidth]{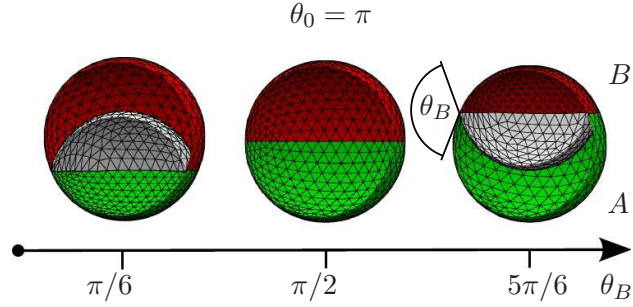}
  \put(95,-4){$\theta_B$}
  \put(12,-4){$\pi/6$}
  \put(45,-4){$\pi/2$}
  \put(80,-4){$5\pi/6$}
 \end{overpic}
\psfragscanoff
 \caption{Configurations of the droplets for various $\theta_B$ in the limiting case $\gamma_{0B}\rightarrow\gamma_{0A}$ and $\gamma_{0A}/\gamma_{AB}\rightarrow \infty$, which corresponds to $\theta_0=\pi$.}
 \label{fig:janus}
\end{figure}

\end{itemize}

In the case of partial-engulfing the curvature $R^{-1}$ of the $AB$-interface can be either positive or negative depending on the contact angles $\theta_A$, $\theta_B$ and the volume ratio $V_B/V_A$. Let us consider the pressure balance in the case of arbitrary $\theta_A$ and $\theta_B$, which corresponds to the geometry depicted in Fig.\ \ref{fig:diagram}. Note that the Laplace pressure inside the phase $A$ due to the curvature $R_A^{-1}$ and the Laplace pressure inside the phase $B$ due to the curvature $R_B^{-1}$ differ by the Laplace pressure due to the curvature $R^{-1}$ of the $AB$-interface (the sign of the curvature is chosen as in Fig.\ \ref{fig:diagram}). This relation can be quantified as
\begin{equation}
 \dfrac{\gamma_{0B}}{R_B}-\dfrac{\gamma_{0A}}{R_A}=\dfrac{\gamma_{AB}}{R}. \label{eq:Laplace1}
\end{equation}
Furthermore, the surface tensions $\gamma_{0A}$ and $\gamma_{0B}$ can be expressed via the contact angles $\theta_A$ and $\theta_B$ by using the law of sines for the Neumann's triangle (see Fig.\ \ref{fig:diagram}), which reads $\gamma_{0B}/\sin\theta_B=\gamma_{0A}/\sin\theta_A=\gamma_{AB}/\sin(\pi-\theta_A-\theta_B)$. Together with Eq.\ (\ref{eq:Laplace1}) this leads to:
\begin{equation}
 R=\sin(\theta_A+\theta_B)\left(\dfrac{\sin\theta_B}{R_B}-\dfrac{\sin\theta_A}{R_A}\right)^{-1}. \label{eq:radius}
\end{equation}
We note that the expression in brackets is a difference of two positive terms (because $R_{A}>0$, $R_B>0$, $0<\theta_A<\pi$ and $0<\theta_B<\pi$). Therefore, one can always choose one of the volumes big enough, such that $R_A\rightarrow\infty$ or $R_B\rightarrow\infty$ which can make $R$ either positive or negative, respectively.

\subsection{Janus-like states}\label{sec:janus}
\begin{figure}[ht]
\psfragscanon
 \psfrag{t}[c][c][1]{$\theta_B$}
 \psfrag{v}[c][c][1]{\rotatebox{90}{$(V_B/V_A)^{1/3}$}}
 \psfrag{va}[c][c][1]{$V_A$}
 \psfrag{vb}[c][c][1]{$V_B$}
 \psfrag{r+}[c][c][1]{$R>0,\beta<\theta_B$}
 \psfrag{r-}[r][r][1]{$R<0,\beta>\theta_B$}
 \psfrag{r}[c][c][1]{$R$}
 \psfrag{b}[c][c][1]{$\beta$}
 \psfrag{A}[c][c][1]{$A$}
 \psfrag{B}[c][c][1]{$B$}
 \begin{overpic}[width=0.6\textwidth]{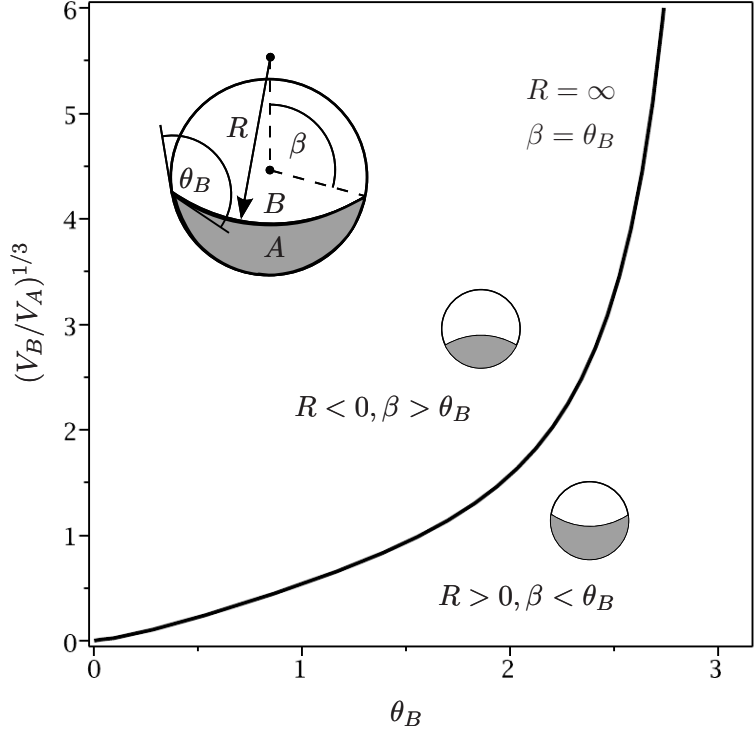}
  \put(68,84){$R=\infty$}
  \put(68,78){$\beta=\theta_B$}
 \end{overpic}
\psfragscanoff
 \caption{Diagram representing the regions of positive and negative curvature $R^{-1}$ of the $AB$-interface in the case of the Janus states. The solid line corresponds to $V_B/V_A=(V_B/V_A)_{crit}$ being the solution of Eq.\ (\ref{eq:ratio}).}
 \label{fig:janus2}
\end{figure}

In the following we focus on the special case of the Janus states, in which $\theta_A+\theta_B=\pi$ and $R_A=R_B$. Apparently, for such a choice the right hand side of Eq.\ (\ref{eq:radius}) is ill-defined. Instead, another geometric property should be used. With the help of an auxiliary variable $\beta$ describing the angular position of the three-phase contact line (see Fig.\ \ref{fig:janus2}) one finds that
\begin{equation}
 R=R_A\dfrac{\sin\beta}{\sin(\theta_B-\beta)}\qquad(\text{Janus}),
\end{equation}
from which it follows that, in the Janus state, the sign of $R$ equals the sign of $\theta_B-\beta$. Accordingly, for $\beta<\theta_B$ one has $R>0$, whereas for $\beta>\theta_B$ one has $R<0$. The limiting case $\beta=\theta_B$ leads to $R=\infty$ which corresponds to a flat $AB$-interface, when both phases form truncated spheres. In such a case the critical volume ratio $(V_B/V_A)_{crit}$ can be calculated from the following geometric relation 
\begin{equation}
 \left(V_B(V_A+V_B)^{-1}\right)_{crit}=((V_B/V_A)_{crit}^{-1}+1)^{-1}=(2+\cos\theta_B)\sin^4(\theta_B/2), \label{eq:ratio}
\end{equation}
where the right hand side is the volume of a unit spherical cap characterized by the polar angle $\theta_B$. In Fig.\ \ref{fig:janus2}, instead of $V_B/V_A$, we use as a variable the ratio of radii of two free spherical drops of volumes $V_B$ and $V_A$, which equals $(V_B/V_A)^{1/3}$. The solid line in the graph corresponds to the solution of Eq.\ (\ref{eq:ratio}) with respect to $(V_B/V_A)_{crit}$. We obtain a diagram in which the region above the line corresponds to negative and the region below the line to positive curvatures of the $AB$-interface.

\section{Experiment}

\begin{figure}[ht]
\centering
\psfragscanon
\psfrag{a}[c][c][1]{$(a)$}
\psfrag{b}[c][c][1]{$(b)$}
 \includegraphics[width=0.9\textwidth]{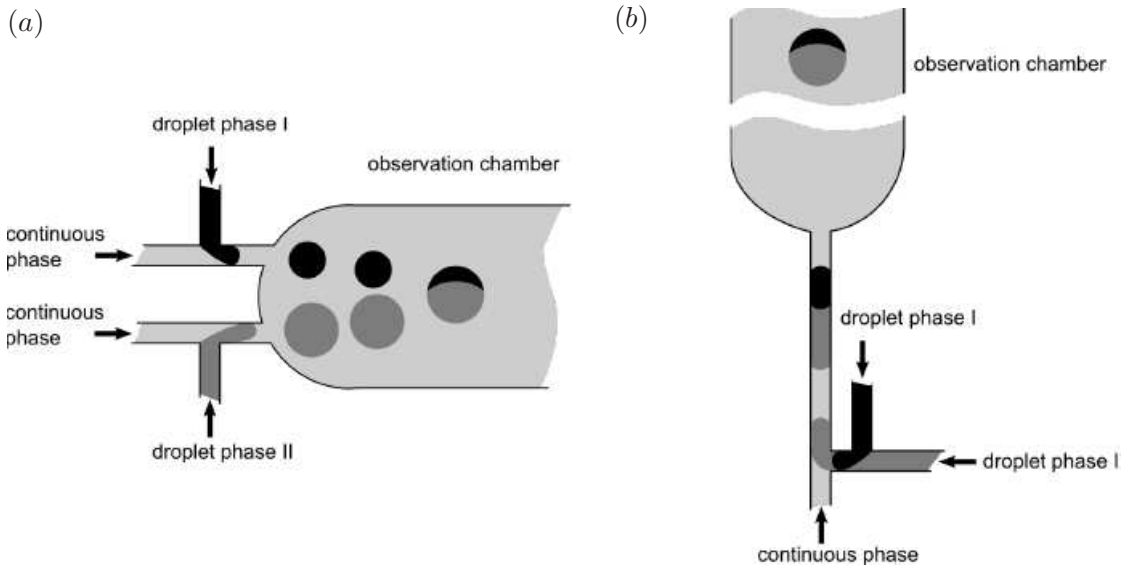}
\psfragscanoff
 \caption{Schematic view of microfluidic chips used to generate and observe double emulsions. Width and hight of channels of the T- junctions are both 200 $\mu m$. The height of the observation chamber is 200 $\mu m$ and the width 3 mm. $(a)$ Two independent T-junctions: droplets merge in the observation chamber. $(b)$ A single two-step T-junction: the double droplets are formed at the second T-junction.}
 \label{fig:exp_setup}
\end{figure}

\begin{figure}[ht]
\psfragscanon
\psfrag{a}[c][c][1]{$(a)$}
\psfrag{b}[c][c][1]{$(b)$}
\psfrag{261}[c][c][1]{$R=261$}
\psfrag{174}[c][c][1]{$R=174$}
\psfrag{155}[c][c][1]{$R=155$}
\psfrag{th}[c][c][1]{$\theta_B$}
\psfrag{e}[c][c][1]{engulfing}
\psfrag{pe}[c][c][1]{partial-engulfing}
\psfrag{ne}[c][c][1]{non-engulfing}
\psfrag{janus}[c][c][1]{partial-engulfing, Janus states, $R_A=R_B=163$}
 \includegraphics[width=0.7\textwidth]{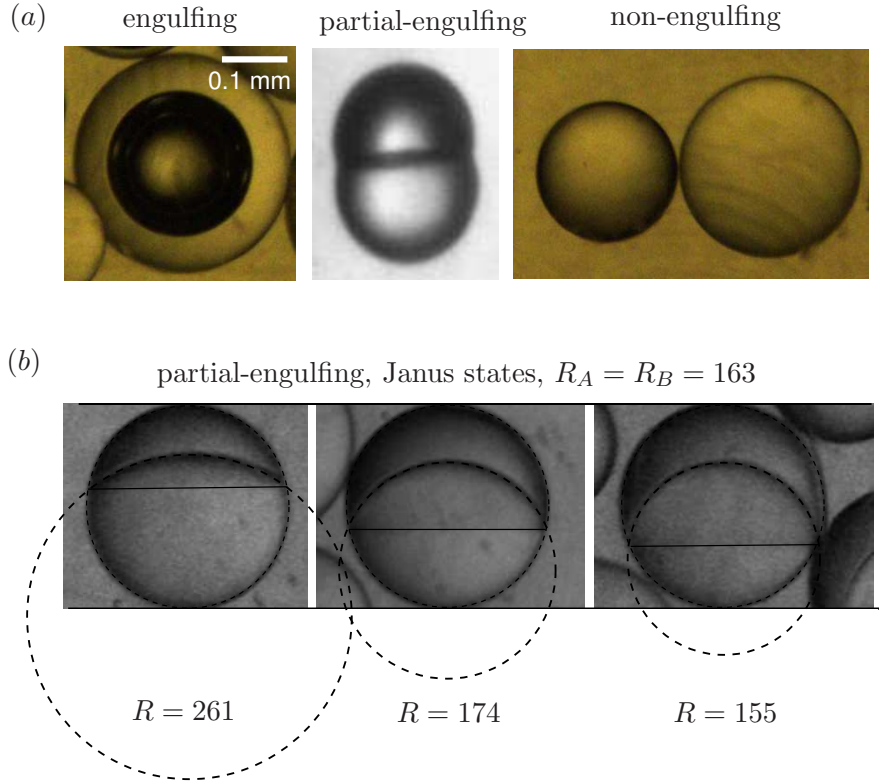}
\psfragscanoff
\caption{$(a)$ The photographs illustrating the three possible equilibrium structures. An air bubble absorbed by a silicone oil droplet corresponds to the engulfing configuration, whereas the droplets of silicone and rapeseed oil in water + 2\% SDS to the partially-engulfing one. The third photograph represents two oil droplets before coalescence (here, a non-engulfing state is not actually the equilibrium one). $(b)$ The dependence of the geometry of the Janus state on the volume ratio. The radius $R$ of curvature (indicated by the dashed circles) of the inner interface decreases with the decreasing volume ratio of lower to upper phase, in qualitative agreement with the theory (Fig.\ \ref{fig:janus2}). The thin horizontal solid lines indicate the three phase contact lines.}
\label{fig:exp_results}
\end{figure}

We produced and observed emulsions in the microfluidic device fabricated  via direct milling in polycarbonate sheets (Macroclear, Bayer, Germany) using a CNC milling machine (MSG4025, Ergwind, Poland). Microfluidic chip consists of two independent T-junctions~\cite{Okushima2004} (see Fig.\ \ref{fig:exp_setup}$(a)$) fed with liquids by the use of four syringe pumps (PHD2000 Harvard Apparatus, USA). The size of droplets is adjusted by the choice of rates of flows of each liquid. Each of the T-junctions produces droplets of different phase. Generated droplets of both phases are directed to the observation chamber where droplets coalesce spontaneously forming a double emulsion. 

Independently we have also used another microfluidic chip consisting of two units: one used for generation of droplets, the other one for observation. In this case the generation part is two step T-junction: at the first junction two streams of droplet phases are joined and than at the second junction both droplet phases are introduced into the continuous phase (see Fig.\ \ref{fig:exp_setup}$(b)$). The width and the hight of the channels is 200 $\mu m$. Generated emulsion is directed to the observation chamber of the width and height $3 mm$ and the length $30 mm$. During experiments the chip is oriented vertically. It allows us to observe freely sedimenting droplets of emulsion flowing along observation chamber due to buoyancy. Owing to the density difference between the liquids, droplets tend to orient their axes of symmetry along the vertical direction. Images of droplets were collected by CCD camera (UI-2230SE, IDS, Germany) connected with stereoscope (SMZ800 with objective 1xWD70, Nikon, Japan). Axis of the visualisation system was placed vertically to observe droplets perpendicular to their axis of symmetry. It is important for the accurate measurement of contact angle. 

We used the droplet on demand system (DOD) to control the size of droplets and ratio of volumes of both droplet phases. The DOD system is based on a standard plunger-type solenoid valve (V165, Sirai, Italy) and was described by Chursky \textit{et al.}~\cite{Churski2010}. Each liquid phase is pressurized in containers and introduced into the chip via steel capillaries (O.D. 400 $\mu m$, I.D. 205 $\mu m$, Mifam, Poland). Connections between containers and inlets of the chip are controlled by valves. It enables to open or close the flow of each phase separately. Owing to the simple linear correspondence between time of valve opening and the volume of introduced liquid we could form droplets of a given ratio of volumes.


\end{document}